\documentclass[11pt,a4paper]{article}

\usepackage{amsmath,amssymb}
\usepackage{epsfig,graphicx}
\usepackage{subfigure}
\usepackage{graphicx}
\usepackage{rotating}
\usepackage{cancel}
\usepackage{bm}
\usepackage{color}
\usepackage{comment}
\usepackage{cite}

\def\beq{\begin{equation}}
\def\eeq{\end{equation}}

\begin{document}
\numberwithin{equation}{section}
\title{{\normalsize  DCPT/11/110; DESY 11-163; IPPP/11/55 \hfill\mbox{}\hfill\mbox{}}\\
\vspace{2.5cm} \Large{\textbf{Prospects for Searching Axion-like Particle Dark Matter with Dipole, Toroidal and Wiggler Magnets
\vspace{0.5cm}}}}

\author{Oliver K. Baker$^1$, Michael Betz$^{2}$, Fritz Caspers$^{2}$, Joerg Jaeckel$^{3}$, Axel Lindner$^{4}$, \\
Andreas Ringwald$^{4}$, Yannis Semertzidis$^{5}$, Pierre Sikivie$^{6}$, Konstantin Zioutas$^{7}$.\\[2ex]
\small{\em $^1$ Department of Physics, Yale University,} \small{\em New Haven, CT 06520-8120, 
United States}\\[0.5ex]
\small{\em $^2$ CERN, CH-1211 Geneva, Switzerland}\\[0.5ex]
\small{\em $^3$Institute for Particle Physics Phenomenology, Durham DH1 3LE, United Kingdom}\\[0.5ex]
\small{\em $^4$Deutsches Elektronen Synchrotron DESY, Notkestrasse 85, D-22607 Hamburg, Germany}\\[0.5ex]
\small{\em $^5$Brookhaven National Laboratory, NY-USA}\\[0.5ex]
\small{\em $^6$Department of Physics, University of Florida, Gainesville, FL 32611, USA }\\[0.5ex]
\small{\em $^7$University of Patras, Patras, Greece}\\[0.5ex]
}
\date{}
\maketitle

\begin{abstract}
\noindent
In this work we consider searches for dark matter made of axions or 
axion-like particles (ALPs) using resonant radio frequency 
cavities inserted into dipole magnets from particle 
accelerators, wiggler magnets developed for accelerator based advanced light sources, 
and toroidal magnets similar to those used in particle 
physics detectors.  We investigate the expected sensitivity 
of such ALP dark matter detectors and discuss the engineering 
aspects of building and tuning them.
Brief mention is also made of even stronger field magnets that are becoming available due to improvements in magnetic technology.
It is concluded that new experiments utilizing already existing magnets could greatly enlarge the mass region in searches for axion-like dark matter particles.
\end{abstract}

\newpage

\section{Introduction}
\begin{figure}[t]
\begin{center}
\includegraphics[width=11.5cm]{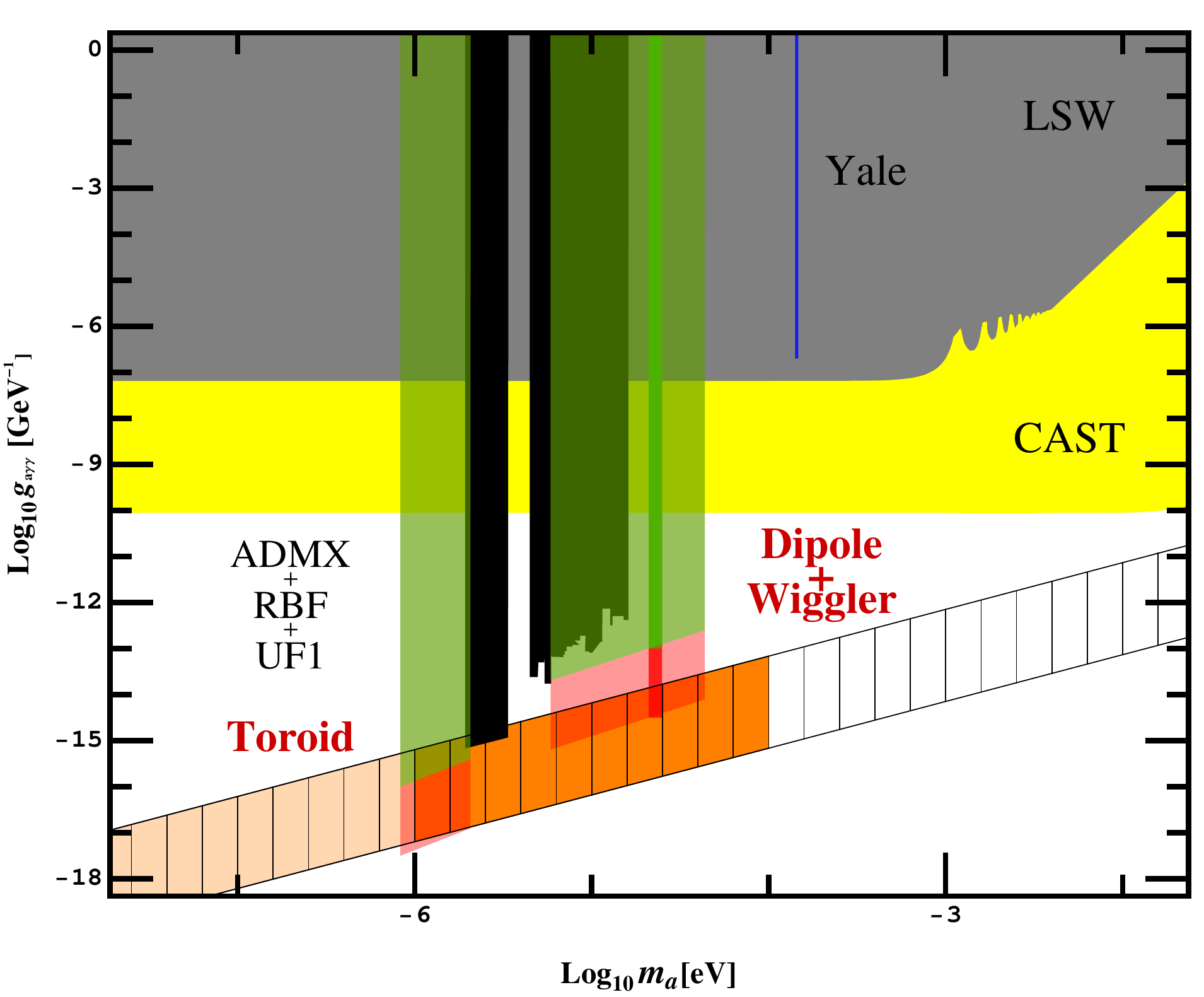}
\caption{Existing bounds for axion-like particles~\cite{DePanfilis:1987dk,Wuensch:1989sa,Hagmann:1990tj,Asztalos:2001tf,Asztalos:2009yp,Ehret:2010mh,Aune:2011rx}. Compilation adapted from~\cite{Jaeckel:2010ni}. 
The true QCD axion-region is marked as the hatched band. In the orange region axions are a natural candidate for dark matter. 
In the lighter shaded orange area axions can still be dark matter but with decreasing mass this requires an increasing amount of finetuning.
Axion-like particles can be dark matter in a large part of the so far untested parameter space (cf.~\cite{Arvanitaki:2009fg,WISPy}).
The right green (red) regions are the conservatively (realistically) expected sensitivity of a dipole or wiggler search for axion-like particle dark matter.
The more solidly shaded part in the middle of this region corresponds to the region probed by a single cavity with parameters as discussed in Sect.~\ref{thin}. The left green/red region shows the masses and couplings that could be probed in a setup with a toroidal magnet.
}
\label{limits}
\end{center}
\end{figure}

Axions are one of the best motivated dark matter (DM) candidates. Arising as consequence of the Peccei-Quinn solution 
to the strong CP problem~\cite{Peccei:1977hh,Weinberg:1977ma,Wilczek:1977pj,Kim:1979if,Dine:1981rt,Shifman:1979if,Zhitnitsky:1980tq}
they are a natural cold dark matter candidate~\cite{Preskill:1982cy,Abbott:1982af,Dine:1982ah} if their mass 
lies within the range of $(1-100)\,\mu{\rm eV}$\footnote{At even lower masses the axion may still be a viable dark matter candidate if one allows for some finetuning or anthropic reasoning~\cite{Hertzberg:2008wr}. This corresponds to the light orange region in Fig.~\ref{limits}.}. This region is marked in orange in Fig.~\ref{limits}. Beyond the explicit axion there is a wide range of so-called axion-like
particles (ALPs) which have similar couplings but with an often increased coupling constant. They could also be good dark matter candidates~\cite{Arvanitaki:2009fg,WISPy}.
Therefore, searches for axion(-like particle) dark matter have great potential to shed light on one of the most important open questions in physics.

The possibility that axions or ALPs (from now on we will include axions whenever we say ALPs) are dark matter 
both allows and requires new search strategies. 
In contrast to searches for Weakly Interacting Massive Particles, scattering experiments are less suitable to search for ALP dark matter due to their extremely low masses. Hence one usually strives for inducing a conversion of ambient dark matter ALPs into detectable particles.
Most of these searches are based on the coupling of ALPs to two photons\footnote{ALPs could also be scalars instead of pseudo-scalars. In this case the coupling would be $-g_{a\gamma\gamma}/4 \,\,\phi_{{\rm ALP}}F^{\mu\nu}F_{\mu\nu}$. For simplicity in this note we focus on the pseudo-scalar case.},
\begin{equation}
\label{interaction}
{\mathcal L}\supset -\frac{1}{4}g_{a\gamma\gamma}\phi_{{\rm ALP}}F^{\mu\nu}\tilde{F}_{\mu\nu}=g_{a\gamma\gamma}\phi_{{\rm ALP}}{\mathbf{E}}\cdot{\mathbf{B}},
\end{equation}
where $g_{a\gamma\gamma}$ is the coupling constant with mass dimension $-1$ and $\phi_{{\rm ALP}}$ denotes the ALP field.

Most prominently, searches employing microwave cavities, so-called haloscopes, have great potential for detecting ALP dark matter in the $\mu$eV mass 
range~\cite{Sikivie:1983ip}.  Indeed a number of searches using haloscopes have already been undertaken~\cite{DePanfilis:1987dk,Wuensch:1989sa,Hagmann:1990tj}, most recently the ADMX experiment~\cite{Asztalos:2001tf,Asztalos:2009yp}. 
The corresponding limits are shown as the black ``fingers'' in Fig.~\ref{limits} as a function of the ALP mass $m_{a}$.
The principle of the haloscope is as follows. Dark matter axions entering a region permeated by a strong 
external magnetic field ${\mathbf{B}}$ will be  converted into photons (which correspond to the electric field part in the interaction 
term \eqref{interaction}) which can then be detected. This conversion probability can be significantly enhanced if it occurs in a cavity resonant to the energy of the ALPs. Since ALPs are typically a cold dark matter candidate their energy is given essentially by their mass. To cover a wide range of axion masses one therefore has to have setups with cavities resonant at a range of different frequencies. 
Proposing experiments for sensitively probing new frequency ranges is one of the goals of the present note.

Up to now all haloscope searches have used solenoid magnets to provide the magnetic field. However, 
at accelerator and advanced light source facilities a range of magnets are available that have not yet been exploited for their use in searches for ALP DM. 
In particular we have the typical dipoles used for the accelerator itself as well as toroidal magnets used in particle detectors. 
In addition we have the wiggler magnets used for  creating synchrotron and free electron laser light.
The aim of this note is to investigate these magnets with respect to their use in haloscope searches for ALP DM.
As we will see these new types of magnets allow to explore new mass ranges as well as having the potential for better sensitivity.
In the following sections \ref{dipole}, \ref{wiggler} and \ref{toroid} we will go through the three mentioned types of magnets, discuss the required cavities and mechanisms for tuning them and provide estimates for the achievable sensitivity.
In Sect.~\ref{future} we will then briefly discuss possibilities for future magnets. Finally we conclude in Sect.~\ref{conclusions}.

\section{Dipole Magnets}\label{dipole}
Let us first consider the use of a typical accelerator type dipole magnet.
Accelerator dipoles offer strong fields in a long region, with the magnetic field oriented perpendicular to the long direction.
Typical parameters are given in Tab.~\ref{table1}.
\begin{table}[t]
\centering
\begin{tabular}{|l||c||c||c||} \hline
Sc-dipole magnet  & $B$ (T)  & $a$ (mm) & L (m)\\
\hline
HERA & 5.5 &  27.5 & 8.8  \\
\hline
LHC  & 9.5  & 25 & 14.3  \\
\hline
Tevatron  & 5 & 24 & 6   \\
\hline
CAST magnet  & 9 & 21.5 & 9.26  \\
\hline
\end{tabular}
\caption{Dimensions of three different types of available superconducting dipole magnets. The bore aperture radius $a$ assumes straightened magnets. (The straightening of HERA dipole magnets is presently under study at DESY in context of the ALPS-II experiment.)}
\label{table1}
\end{table}

\subsection{Sensitivity}\label{thin}
\begin{figure}[t]
\begin{center}
\includegraphics[width=0.4\textwidth]{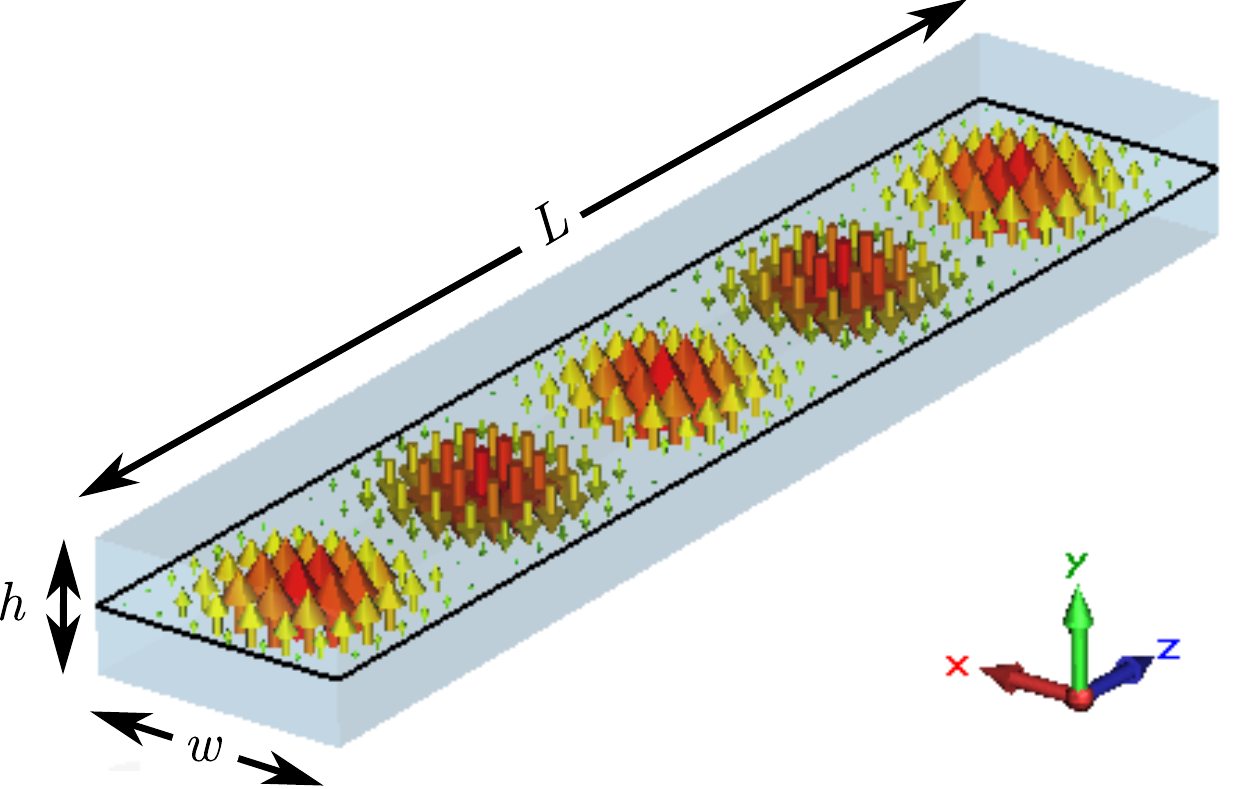}
\caption{Electric field distribution of the TE$_{105}$ mode in a rectangular waveguide.}
\label{wgField}
\end{center}
\end{figure}

A simple setup for a dark matter search could consist of a long thin rectangular cavity which is introduced into the dipole magnet.
We denote by $h$ the side length parallel to the dipole magnetic field, by $w$ the one perpendicular to the dipole magnetic field and by $L$ the length along the dipole magnet (see also Fig.~\ref{wgField}). 

Geometrically the maximal extent of such a cavity is constrained by the aperture $a$ of magnet via
\begin{equation}
w\leq 2 a \sin(\alpha/2),\quad h\leq 2 a \cos(\alpha/2),
\end{equation}
where 
\begin{equation}
\tan(\alpha/2)=\frac{w}{h}.
\end{equation}

For example with an aperture $a\geq20$mm (all magnets in Tab.~\ref{table1} allow for this) and a ratio of width to height $w/h=3/2$
we can fit a cavity of $w\times h=30{\rm mm}\times20{\rm mm}$ into the tube.
For the length we will take $L=8.5{\rm m}$. In the following we will use these numbers for our simple estimate.

For dipole magnets, alignment of the electric field of the cavity with the external magnetic field is achieved by using TE modes.
An exemplary field configuration of a TE$_{105}$ mode is depicted in Fig.~\ref{wgField}.
For a rectangular cavity the frequency of a general TE$_{lmn}$ (where the indices are ordered $w$, $h$, $L$) mode is given by
(in the following we mostly use $c=\hbar=1$ but we will restore these constants from time to time for clarity)
\begin{equation}
\label{equ:resF}
\omega_{lmn}=\sqrt{\left(\frac{l\pi}{w}\right)^2+\left(\frac{m\pi}{h}\right)^2+\left(\frac{n\pi}{L}\right)^2}.
\end{equation}  

A suitable mode is TE$_{101}$.
Indeed for the given cavity geometry this mode is the lowest possible mode. This is advantageous because the lowest
modes are usually the most isolated from other modes.   That is, regions with many overlapping modes can lead to reduction
in sensitivity, so this favors lower modes for the experiment. 
The electric field of a TE$_{l0n}$ mode is oriented along the magnetic field of the dipole and is given by
\begin{equation}
\label{fielddist}
E_{y}\sim \sin\left(\frac{\pi l x}{w}\right)\sin\left(\frac{\pi n z}{L}\right).
\end{equation}
The frequency is
\begin{equation}
\omega_{101}=(2\pi) 5\,{\rm GHz}=2.1\times10^{-5}\,{\rm eV}.
\end{equation}
This frequency (cf. the solid green/red area in Fig.~\ref{limits}) is nicely complementary to the ADMX setup~\cite{Asztalos:2001tf,Asztalos:2003px} (black region in Fig.~\ref{limits}) as well as the 
searches at Yale~\cite{Martin:2011} (blue region in Fig.~\ref{limits}).

Equation~\eqref{fielddist} can be used to calculate the geometry factor relevant for ALP DM searches~\cite{Sikivie:1985yu},
\begin{equation}
\label{equ:geomFactor}
C=\frac{\left(\int dV\, \mathbf{E}_{\rm cav}({\mathbf{x}})\cdot\mathbf{B}_{0}({\mathbf{x}})\right)^2}{V|\mathbf{B}_{0}|^2\int dV\, \epsilon({\mathbf{x}})
\mathbf{E}^{2}_{\rm cav}({\mathbf{x}})},
\end{equation}
where $\mathbf{E}_{\rm cav}$ is the electric field of the cavity (with volume $V$) and $\mathbf{B}_{0}$ is the external magnetic field provided by the dipole. Moreover, $\epsilon({\mathbf{x}})$ is the dielectric constant. When using dielectric inserts, as discussed in Sect.~\ref{tune}, 
$\epsilon$ may depend on the position ${\mathbf{x}}$ inside the cavity. 

For the cavity in question we find
\begin{equation}
\label{geometry}
C_{10n}=\frac{64}{\pi^4}\frac{1}{n^2}=\frac{0.66}{n^2}\quad{\rm for}\,\, n\,\,{\rm odd}.
\end{equation}
For all even $n$ the geometry factor vanishes.
In particular for $n=1$ the geometry factor is $0.66$ and therefore quite sizable.

Let us now estimate the sensitivity of such a setup.
The output power of the cavity is given by~\cite{Sikivie:1985yu} (see also, e.g.~\cite{Asztalos:2001tf}),
\begin{eqnarray}
\label{power}
P_{\rm out}\!\!&=&\!\! g^{2}_{a\gamma\gamma}V B^2\rho_{a} C\frac{1}{m_{a}}\min[Q,Q_{a}]
\\\nonumber
\!\!&=&\!\!1.1\times10^{-26}{\rm W}\left(g_{a\gamma\gamma} 10^{15}{\rm GeV}\right)^2\left(\frac{C}{0.66}\right)
\left(\frac{B}{5{\rm T}}\right)^2\left(\frac{V}{5\ell}\right)\\\nonumber
&&\quad\quad\quad\quad\quad\quad\quad\quad\quad\quad\quad\quad\quad\quad\quad\times \left(\frac{\rho_{a}}{300 {\rm MeV}/{\rm cm}^3}\right)\left(\frac{2.1\times 10^{-5}{\rm eV}}{m_{a}}\right)\left(\frac{Q}{10^3}\right),
\end{eqnarray}
where $Q$ is the loaded\footnote{``Loaded'' in this context means that it is the Q-factor of the cavity coupled to the detector. Using an impedance matched cavity to optimize the power reaching the detector, one has $Q=Q_{0}/2$ where $Q_{0}$ is the unloaded $Q$-factor of the cavity alone.}
$Q$-factor and $Q_{a}\sim 10^6$ is the ALP dark matter $Q$-factor resulting from the energy spread of the ALPs (see Eq.~\eqref{ALPQ} below).
In the second part of the equation we have assumed $Q<Q_{a}$ as a further increase in $Q$ does not increase the power.

The essential question is now if one can detect the resulting small powers.
The time $t$ needed to detect a given amount of power $P$ with a signal to noise ratio $SNR$ is determined by the Dicke radiometer equation,
\begin{equation}
SNR=\frac{P}{k_{B} T_{n}}\sqrt{\frac{t}{b}},
\end{equation}
where $T_{n}$ is the total noise temperature of the receiver and amplifier chain and $b$ is the bandwidth in which the signal power is contained.
In the ALP case the bandwidth of the signal is determined by the energy spread of the dark matter axions which in turn is given by their velocity distribution,
\begin{equation}
\label{ALPQ}
2\pi b_{a}=\frac{m_{a}}{Q_{a}}=m_{a}\left(\frac{1}{\sqrt{1-\Delta v_{a}^2}}-1\right)\sim 10^{-6}m_{a}.
\end{equation}
In the last step we have used that the typical velocitiy spread of dark matter ALPs is expected to be $\Delta v_{a}\sim 300\,{\rm km}\sim 10^{-3} c$.
Therefore the time needed for a single measurement is,
\begin{eqnarray}
\label{time}
t\!\!&=&\!\!(SNR)^2\left(\frac{k_{B} T_{n}}{P_{out}}\right)^2 b_{a}=(SNR)^2\left(\frac{k_{B} T_{n}}{P_{out}}\right)^2\frac{f}{Q_{a}}
\\\nonumber
\!\!&\sim&\!\! 1.1 \times10^{12}\,{\rm s} \left(\frac{SNR}{4}\right)^2\left(\frac{T_{n}}{3{\rm K}}\right)^2 
\left(\frac{B}{5{\rm T}}\right)^{-4}\left(\frac{V}{5\ell}\right)^{-2}\left(\frac{C}{0.66}\right)^{-2}\\\nonumber
&&\quad\quad\quad\quad\,\,
\times 
\left(g_{a\gamma\gamma} 10^{15}{\rm GeV}\right)^{-4}
\left(\frac{\rho_{a}}{300 {\rm MeV}/{\rm cm}^3}\right)^{-2}\left(\frac{10^6}{Q_{a}}\right)\left(\frac{2.1\times10^{-5}eV}{m_{a}}\right)^{-3}\left(\frac{Q}{10^3}\right)^{-2}. 
\end{eqnarray}
With our benchmark cavity and using a quite conservative $Q=1000$, $T_{n}=3{\rm K}$ as well as $g_{a\gamma\gamma}=10^{-15}{\rm GeV}^{-1}$ we find that it would
be necessary to measure for about 35000 years which is clearly unacceptable. 
On the other hand for $g_{a\gamma\gamma}=10^{-13}{\rm GeV}^{-1}$
this immediately reduces to  $11000\, {\rm s}\sim 3\, {\rm hours}$ which is reasonable.
As can be seen from the green finger in Fig.~\ref{limits}, at a frequency of $5\,{\rm GHz}$, corresponding to a mass of $m_{a}=2.1\times10^{-5}{\rm eV}$, couplings of $10^{-13}{\rm GeV}^{-1}$ are deep in the unexplored region. Therefore with such a setup we definitely have an interesting sensitivity for ALP DM~\cite{WISPy}.
To achieve sensitivity to the QCD axion we need to reach couplings of the order of $10^{-15}{\rm GeV}^{-1}$. For this we need to improve some of the experimental parameters. We will return to these improvements shortly.

As the ALP mass is a priori unknown, an experiment should not only be sensitive to one mass
but instead we would like to be able to scan a range of masses. 
Therefore it is important to determine the speed with which such a scan can be performed.
Having determined the time necessary to perform a single measurement it is relatively straightforward to determine the scanning speed.
There is only one thing to note. 
Given that the bandwidth of the axion signal is smaller than the width of the cavity we can perform a simultaneous measurement
of all potential ALP masses within the width of the cavity by doing a fast Fourier transform (FFT) of the output signal.
Accordingly the scanning speed is determined by~\cite{Hagmann:1989hu,Asztalos:2001tf},
\begin{eqnarray}
\label{scanning}
\frac{df}{dt}\!\!&=&\!\! \frac{f}{Q}\frac{1}{t}
\\\nonumber
\!\!&\sim&\!\! \frac{140{\rm Hz}}{{\rm year}}(g_{a\gamma\gamma}\,10^{15}\,{\rm GeV})^4\left(\frac{4}{SNR}\right)
\left(\frac{V}{5\ell}\right)^2
\left(\frac{B}{5{\rm T}}\right)^4\left(\frac{C}{0.66}\right)^2\left(\frac{3{\rm K}}{T_{n}}\right)^2\left(\frac{5{\rm GHz}}{f}\right)^2
\left(\frac{Q}{10^3}\right).
\end{eqnarray}

Again we see that with our benchmark cavity and the assumed values of the thermal noise and magnetic field strength, the scanning speed is very low for couplings $g_{a\gamma\gamma}=10^{-15}\,{\rm GeV}^{-1}$, whereas it is in 
the desired GHz/year range for $g_{a\gamma\gamma}=10^{-13}\,{\rm GeV}^{-1}$.
This further demonstrates that this setup already has pretty good sensitivity for ALP DM.

Let us now briefly consider improved parameter values that still seem to be within reach of current technology (for more details see also the following sections). The first step is to improve
the detector and amplifiers. Noise temperatures close to the quantum limit have been achieved~\cite{Asztalos:2009yp}. For 5 GHz
the quantum noise temperature is $\omega/k_{B}\sim 240$~mK. Therefore let us assume that we can achieve $T_{n}=0.3$~K.
A $Q\sim 2\times10^4$ is also certainly realistic and finally an improvement in the magnetic field to $10$~T is certainly doable.
Using 
\begin{eqnarray}
T_{n}&\rightarrow& 0.3\,{\rm K}\\\nonumber
Q&\rightarrow& 2\times 10^4\\\nonumber
B&\rightarrow& 10 \,{\rm T},\\\nonumber
\end{eqnarray}
gives us a scanning speed of 0.5 GHz/year for a coupling of $10^{-14.5}\,{\rm GeV}^{-1}$ which is within the QCD-axion band. This is shown as the red tip of the green finger in Fig.~\ref{limits}.

In Fig.~\ref{limits} we also show the extended mass range that could be covered by constructing a series of 4 cavities each (down-)tunable by $20\%$ (see Sect.~\ref{tune} below) and with frequencies $6, 7.2, 8.6,10.4$~GHz while scaling the volume and quality factors appropriately.
This corresponds to the lighter shaded area to the right of the green/red finger.

\subsection{Engineering aspects}
Let us now turn to the practical engineering aspects and limits of this kind of long and thin microwave resonator. As we increase the aspect ratio $L/w$ of the structure, we expect two problems to appear: the $Q$-value will decrease and the resonant frequencies of the modes will converge to a single value. In both cases the sensitivity with respect to detecting ALPs will decrease. Let us now investigate how severe those effects are.

The unloaded $Q$-factor, $Q_{0}$,  of the TE$_{101}$ mode in a rectangular cavity as a function of its geometry is given 
by,
\begin{equation}
\label{equ:qCav}
Q_0 = \frac{2 \pi c}{\omega_{101}\delta } \cdot \frac{h (w^2 + L^2)^{3/2}}{2 L^3 (w + 2 h) + w^3 (L + 2 h) },
\quad \quad
\delta = \sqrt{\frac{2}{\omega_{101} \sigma \mu}},
\end{equation}
where $\delta$ is the skin-depth of the cavity walls, $\mu$ is the permeability and $\sigma$ the conductivity of the metal \cite{src:meinkeGrundlach}.
In the following we use $\delta=10^{-6}\,{\rm m}$ for copper at the resonant frequency of our example setup.

Note that $\delta$ as given in Eq.~\eqref{equ:qCav} is an approximation only valid at room temperature.
At cryogenic temperatures (T $\leq 20^{\circ}$ K) and high frequencies (f $\geq$ 1 GHz), the mean free path of electrons in the copper becomes comparable with or greater than the classical skin depth given by Eq.~\eqref{equ:qCav}. Then one enters the anomalous skin depth regime where $\delta \propto \omega^{-1/3}$ \cite{src:aSkin}.
Hence significantly better values for $\delta$ can be achieved at lower temperatures\footnote{
Note that the DC conductivity of metal is increased at cryogenic temperatures by the residual resistance ratio (RRR), values of $\approx$ 200 are realistic for pure copper \cite{src:RRR}. 
Nonetheless strong magnetic fields can influence charge carriers in the metal (Magnetoresistance), therefore reducing its conductivity by a factor of $\approx$ 15  \cite{src:magnetoresistance}.}.

The $Q$-factor reduces with length $L$, however it approaches a certain ``worst case'' value for very long structures ($L > 20 w$), which is given 
by,
\begin{equation}
\label{equ:qCavLong}
Q_0 = \frac{2 \pi c}{\omega_{101} \delta} \cdot \frac{h}{2 w + 4h}.
\end{equation}

We can compare the analytical equations to numerical data from the electromagnetic simulation software CST microwave studio and the results agree within an error of 0.2~\%.

In our example ($w=30\,{\rm mm}$; $h = 20\,{\rm mm}$) the worst case $Q$-factor for very long structures is $Q_0$ = 11200 which is only 16 \% below the one for short structures.

A deeper understanding of the decrease in $Q$ value can be gained by viewing the structure as waveguide. In the limit of an infinitely long geometry and with the condition of having only one half wave along the length, the waveguide needs to be operated exactly at its cut-off frequency. It shall be made clear that the waveguide attenuation at its cut-off frequency is not infinite as many approximated formulas in microwave engineering literature suggest. The increase in attenuation reflects the 16 \% decrease of $Q$-factor in our example. The exact derivation for the attenuation near cut-off is given in \cite{src:waveguideCutoff}.

The second problem, limiting the length of the cavity is the spacing in frequency to the next higher order mode. If the aspect ratio $L/w$ becomes large, Eq.~\eqref{equ:resF} predicts that the modes converge towards a common resonant frequency. This is unfavorable as two modes with similar resonant frequency will couple to each other and exchange energy. They do this because of resistive losses in the wall currents. 
This is different from the case of a lossless waveguide where the modes are completely orthogonal to each other.

The energy of the unwanted mode is dissipated in the structure and lost because it is not coupled out and guided towards the signal processing chain. Thus we demand the minimum spacing between the resonant frequency of 2 modes to be equal to their 6 dB bandwidth 
Eq.~\eqref{equ:6dBBW}. Then only up to 25 \% (6 dB) of the signal power might be lost through mode coupling. Using this definition we can derive a boundary for the maximum cavity length.

The resonant frequency increases monotonically with the order of a mode, so if we determine the distance in frequency between our working mode (TE$_{101}$) and the next higher order one (TE$_{102}$) we can assume that all other modes are further spaced away. 
In general, the distance between the two center frequencies is given by, 
\begin{equation}
\label{equ:modeSpacing}
\Delta \omega = \omega_{10n}-\omega_{10,n-1}\approx \omega_{101} \cdot \frac{2n-1}{2} \cdot \left(\frac{w}{L}\right)^2.
\end{equation}
Assuming as a first approximation that the $Q$-factors of two neighboring modes are equal,
the 6 dB bandwidth of a TE$_{10n}$ mode is given 
by 
\begin{equation}
\label{equ:6dBBW}
BW_{6 dB} = \frac{2~\omega_{10n}}{2\pi Q} \approx \frac{4\omega_{101}}{2\pi Q_{\mathrm{0}}},  
\end{equation}
where we have assumed $Q_{0}\approx 2 Q$ and $\omega_{10n}\approx\omega_{101}$ in lowest order for small to moderate $n$.
Combining these equations and solving for L we get a maximum length,
\begin{equation}
\label{equ:maxLength}
L_{\rm max}=w\sqrt{\frac{(2n-1)\pi Q_{0}}{4}}=1.5 {\rm m}\left(\frac{w}{3 {\rm cm}}\right)\left(\frac{2n-1}{3}\right)^{1/2}\left(\frac{Q_{0}}{1000}\right)^{1/2} .
\end{equation}

For our benchmark cavity in the TE$_{101}$ mode this gives us an allowed length of about 1.5~m.
As it seems the issue of mode crossing seems to prevent us from making use of a single long cavity over the full length of the dipole.
Note, however, that the allowed length is proportional to $\sqrt{Q}$. Therefore already with a $Q\sim 20000$ this problem basically goes away for the typical dipole magnet.
Moreover, as we will seen in Sect.~\ref{sectioned} we always have the option to use an array of several cavities with external power combiners. Our result predicts that only a few cavities are needed to fill a typical magnet, making this approach technically feasible.

\begin{figure}[t]
\begin{center}
\includegraphics[width=1.0\textwidth]{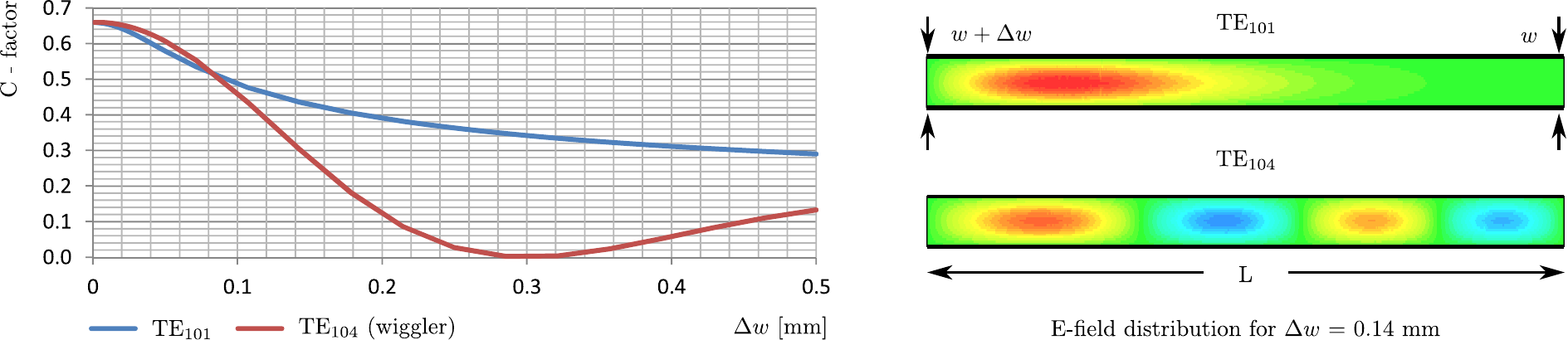}
\caption{Left: Geometric form factor according to Eq.~\ref{equ:geomFactor} as a function of $\Delta w$. Right: Snapshot of the electric field in the cavity, showing mode localization for $\Delta w$ = 0.14 mm. The dimensions of the cavity are w=30, h=20, l=1500 mm. Note that the field plots are not to scale.}
\label{fig:localization}
\end{center}
\end{figure}
Another engineering aspect, giving us upper limits on the mechanical manufacturing tolerances of the resonator is mode localization~\cite{Hagmann:1989hu}. The effect can be observed, if the width of the cavity does not stay constant along its length. A good model to investigate the sensitivity to slight changes of geometry is a trapezoidal structure: going from one end to the other in longitudinal direction, the width of the cross section increases from $w$ to $w + \Delta w$.
The field inside tends to localize in the wider part of the volume and falls off exponentially towards the narrower part, decaying faster with larger values of $\Delta w$. This effect reduces the geometry factor C, as the volume containing no field is not sensitive to converted ALPs.

To investigate the impact of mechanical tolerances on the C-factor, a numerical simulation of a 1,5 m long cavity has been carried out. The resulting C-factors, as a function of $\Delta w$ are shown in Fig.~\ref{fig:localization}. They have been calculated for a single TE$_{101}$ mode cavity in a dipole magnet as discussed in this section and for a wiggler type TE$_{104}$ configuration with 4 magnetic poles as discussed in Sect.~\ref{ch:wiggler}. 
As can be seen from the figure tolerances $\sim 0.1\,{\rm  mm}$ in the transversal direction at one end of the cavity is leads to are acceptable as they only lead to an order 20\% decrease in the geometry factor. For the dipole case even larger deviations from the ideal shape $\sim 0.5\,{\rm mm}$ reduce the geometry factor by not much more than 50\%. This is a worst case scenario because the simulation code does not consider the effect of metallic losses in the waveguide walls and assumes an infinite $Q$. For a lossy structure, the cavity will be less sensitive to mechanical tolerances.

\subsection{Tuning methods}\label{tune}
Finally let us also briefly comment on possible tuning mechanisms.
One way to tune this type of cavity could be to simply move one of the side walls of the cavity.
(In contrast moving the top and bottom lids of the cylindrical cavity used in ADMX does not change the frequency.)
However the sliding contacts needed for such a setup are very difficult to design and prone to failure 
in a cryogenic vacuum enviroment. Therefore this approach is feasible only for quite small $Q$-values.

An alternative option is to place dielectric material inside the cavity \cite{src:axionAntennas}. Depending on its volume, position and
shape, it will change the mode structure and resonant frequency. Note that for a partially filled volume, pure TE and
TM modes can not propagate. Instead we will observe hybrid modes with weak field components in longitudinal direction.

\begin{figure}[t]
\begin{center}
\includegraphics[width=1.0\textwidth]{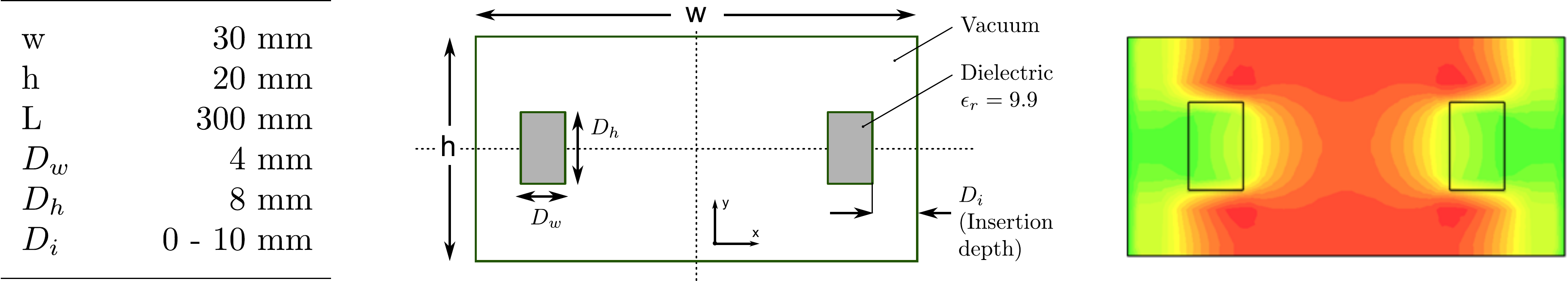}
\caption{Cross section of the waveguide with dielectric tuning rods on each side. The relative electric field intensity in $y$-direction (parallel to the static magnetic field of the external magnet) is shown in color code (blue = negative, green = neutral, red = positive).}
\label{fig:wgCrossSection}
\end{center}
\end{figure}

We propose two movable rods of Alumina along the sides of the cavity structure as indicated in the cut-away view of the waveguide's cross section in
Fig.~\ref{fig:wgCrossSection}.

A straightforward mechanical system makes them move towards the middle of the cavity. Electromagnetic waves inside the dielectric material propagate with a lower phase velocity, which changes the dispersion relations and the field in the cavity. The closer the dielectric rods are to the center of the cavity, the lower its resonant frequency. The sensitivity of the axion antenna is shifted towards ALPs with lower masses.

\begin{figure}[t]
\begin{center}
\includegraphics[width=\textwidth]{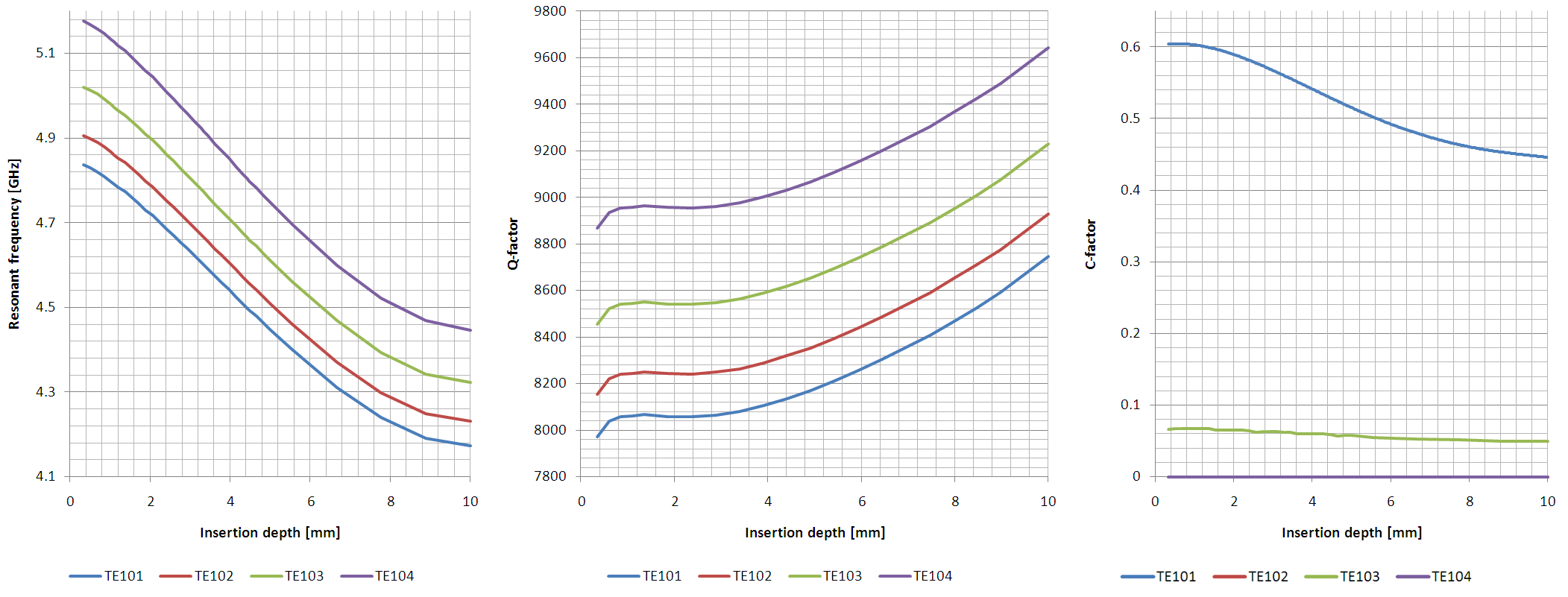}
\caption{Result of a numerical simulation of the tunable waveguide structure. From left to right: Resonant frequency, unloaded $Q$-factor and the geometric C-factor are plotted vs. the insertion depth of the dielectric rods.}
\label{fig:diagTune}
\end{center}
\end{figure}

The results of a numerical simulation with CST microwave studio
can be seen in Fig.~\ref{fig:diagTune}. The geometry provides a 13 \% adjustment range over $f_{\mathrm{res}}$ while keeping the electric field distribution fairly
homogeneous over a big volume. 

For big insertion depths ($D_i > 4\,{\rm mm}$), a significant part of the field is guided inside the dielectric rods instead of the waveguide walls. The structure converges towards a dielectric waveguide -- which is the equivalent of an optical monomode fibre in the microwave range. As this kind of structure does not suffer from resistive losses, the Q-factor increases with the dielectric rods positioned closer to the middle. On the other hand, placing dielectric materials in the active volume degrades the geometric factor according to Eq.~\eqref{equ:geomFactor}, so in total there is no gain in detection sensitivity.

The geometric factor C has been calculated numerically by changing the integration over the cavity volume in Eq.~\eqref{equ:geomFactor} to a summation over sample points in 3D space. Implementation within microwave studio as a VBscript
allows to conveniently investigate arbitary cavity geometries. The results depicted in Fig. \ref{fig:diagTune} show a worst-case degradation of the geometric factor of $\approx 16 \%$ through the dielectric tuning rods, which is acceptable considering that the Q-factor increases by the same percentage.

The advantage of tuning with dielectric materials is that no electrical sliding contacts are needed.
Those are extremely difficult to operate reliably in a cryogenic vacuum environment because of bonding,
contact resistance and brittleness issues.

\subsection{Sectioned cavity}\label{sectioned}
One way to avoid the problem of the closely spaced modes is to section the long tube into a set of $N$ separate cavities, each with its individual coupling antenna. The phase matched output signals of the $N$ antennas are fed to an external power combiner.

As the geometry factor for the cavity and mode used above is independent of the length $L$ the geometry factor and therefore the approximate sensitivity is the same as for the previous single cavity.
The only difference is a slightly higher frequency, due to the shorter length $L$. 

Combining $\sim 10$ cavities in this way should be feasible. With some effort perhaps even combining a few 100 might be doable.
This allows for a significant improvement in sensitivity. Already 10 connected cavities improve the sensitivity for the coupling 
$g_{a\gamma\gamma}$ by a factor $10^{0.5}$.  With the more advanced setup discussed at the end of Sect.~\ref{thin} this brings us even deeper into the regime of the true QCD axion.

From the magnet point of view using multiple cavities with a combined length greater than one dipole requires the operation of several magnets.
Although naively one would assume that the costs for this would scale with the number of magnets. However, the cooling and current supply for superconducting magnet can often be connected such that it is not much more costly to operate a significant number of magnets than it is to operate a single one. Indeed, e.g. in the HERA tunnel one already has sections with 52 magnets connected in this way.
These magnets are not (yet) straightened. This is not necessarily a big problem as it is possible to slightly bend the cavity which should not change its properties dramatically. Alternatively one could go to slightly smaller cavities with correspondingly higher frequencies.

However, there is a physical limit to the number of cavities one can align along a line.
The coherence length of ALP DM matter particles is the maximal length over which the axion field can be thought to be in phase.
It is given by the width of the momentum distribution\footnote{If a significant part of the DM density in our neighborhood 
originates from a late infall of dark matter~\cite{Sikivie:2001fg} this part would have a narrower velocity distribution. Accordingly in this case the coherence length would be much longer.}
$\Delta p_{a}$,
\begin{equation}
l_{\rm coherence}\sim \frac{1}{\Delta p_{a}}
=\frac{1}{m_{a}\Delta v_{a}}\sim 10\,{\rm m}\left(\frac{2.1\times10^{-5}{\rm eV}}{m_{a}}\right)\left(\frac{10^{-3}}{\Delta v_{a}}\right). 
\end{equation}
In the frequency range we are interested in this severely limits the number of cavities which can be used if they are arranged one after the other in a row. On the other hand this does not prevent us from using a larger number of magnets (e.g. 50 or so HERA magnets), it only means that they
should not be arranged in a single line but in a more compact arrangement that stays within the boundaries set by the coherence length 
(e.g. a square grid). 

In principle one can also combine cavities over longer length than the coherence length. However, incoherent addition of the power only leads to an increase of the output power by the square root of the number of cavities instead of a linear increase in the coherent case. Accordingly the sensitivity
then only improves by a fourth root of the number of cavities.

\section{Wiggler}\label{wiggler}
Wiggler and undulator magnets developed to generate synchrotron or free electron laser light from electron beams also provide
interesting possibilities for searching ALP DM.
A selection of magnets together with their relevant parameters is given in Tab.~\ref{table2}.

\begin{table}
\centering
\begin{tabular}{|l||c||c||c||c||} \hline
Wiggler magnet  & $B$ (T)  & bore $w\times h$ (mm$^2$) & $d$=domain size (mm) & length (mm)\\
\hline
SPRING-8 & 10 &  100$\times$20& 333& 1000  \\
\hline
SC-20  & 3.5  & 20$\times$ 8& 90&1800 \\
\hline
BESSY-2 & 7 & 110$\times$13 &148 & 1924  \\
\hline
\end{tabular}
\caption{{\footnotesize{Dimensions of three different exemplary types of available wiggler magnets. }}}
\label{table2}
\end{table}

\subsection{Single long cavity in higher mode}
\label{ch:wiggler}
Let us take the SPRING-8 magnet as an example. From its width and height our benchmark cavity also fits into this magnet.
If only a single wiggler of this type is available the useable cavity length, however, is reduced to $\sim$1~m. However, there is absolutely no fundamental problem in combining a sizeable number of wigglers to achieve a total cavity length of about 10~m.

Let us now take a look at the required modes and the achievable sensitivity.
To make optimal use of the cavity we use a TE$_{10n}$ mode where $n$ is given by the number of alternating field regions. For the
SPRING-8 magnet $n=3$.

As long as we choose $n$ to be equal to the number of magnetic domains in the wiggler the geometry factor is given by Eq.~\eqref{geometry}
for the (101)-mode, $C_{101}=0.66$.

The power output, measuring time and scanning speed are given by Eqs.~\eqref{power}-\eqref{time}.
With the same assumptions for the detector and cavity quality factor the sensitivity for a single wiggler is reduced by a factor three compared to the (longer) dipole. Combining several wigglers to achieve the same total length the sensitivities are equal.

An advantage of the wiggler configuration is that the mode spacing Eq.~\eqref{equ:modeSpacing} grows with $n$, i.e. the number of oscillations
in the wiggler. The problem of overlapping modes is therefore somewhat reduced.
Using this we find for the maximal useable length without sectioning,
\begin{equation}
\label{maxlength2}
L_{\rm max}\approx d\frac{\pi Q_{0}}{2}\frac{w^2}{d^2}= 4.7 \,{\rm m}\left(\frac{30 \,{\rm cm}}{d}\right)\left(\frac{w}{3 \,{\rm cm}}\right)^2\left(\frac{Q_{0}}{10^3}\right),
\end{equation}
where $d$ is the domain size of the wiggler magnet.
We can see that using a wiggler magnet significantly reduces the need for sectioning.

The bigger width of many of the wiggler magnets could also be used to probe lower frequencies. 
Using the full width of 10 cm of the SPRING-8 magnet reduces the frequency to about 1.5 GHz, while only increasing the sensitivity due to the additional volume while at the same time benefitting from the increased ALP number density. In addition one can hope for an improved $Q$ 
according to Eq.~\eqref{equ:qCav}.
Building a set of tunable cavities such that the whole frequency range $1.5-10$~GHz is covered, one could hope to scan the lightly shaded region shown in Fig.~\ref{limits}. 

\subsection{Cavity with 180 degree twists}
For a rectangular cavity in a dipole magnet all TE$_{10n}$ modes with $n>1$ suffer from a low geometric form factor C. This is predicted by 
Eq.~\eqref{equ:geomFactor} and a direct cause of the sinusoidal electric field distribution $E_y$ along the length of the cavity given by Eq.~\eqref{fielddist}. For $n>1$ there are positive and negative values for $E_y$ in different parts of the cavity, cancelling each other out partly or completely in Eq.~\eqref{equ:geomFactor} and leading to a small geometric form factor.

\begin{figure}[t]
\begin{center}
\includegraphics[width=0.4\textwidth]{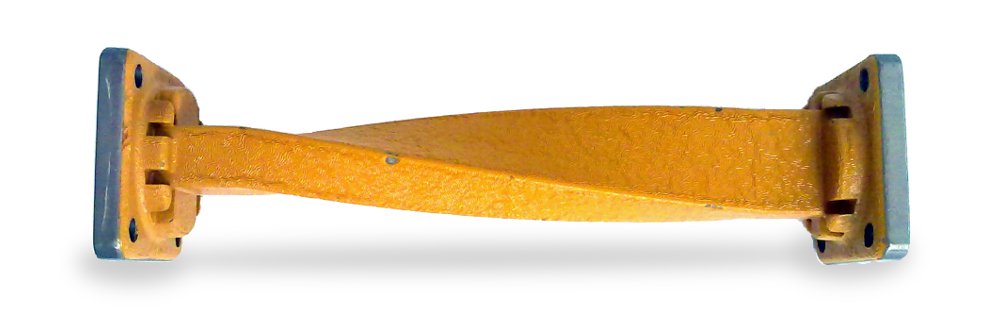}
\caption{Twisted waveguide section (90 degree). A structure consisting of several straight sections connected by 180 degree twists is predicted to have a high geometric factor close to $C\approx 0.6$ in a dipole magnet while avoiding many engineering challenges of other setups.}
\label{fig:twistedWG}
\end{center}
\end{figure}

One way to work around this issue is proposed in Sect.~\ref{ch:wiggler} and involves changing the polarity of the external magnetic field in those regions, where the electric field points in the unwanted direction, thus recovering the geometric form factor close to its theoretical maximum of $C \approx 0.66$.
Alternatively the cavity can be mechanically rotated, changing the polarity of the electric field inside. We propose an alternation of a straight rectangular piece and a special element, smoothly twisting the structure by 180 degree around its axis. This alternation is repeated to fill up the dipole magnet. A similar, commercially available waveguide twist for 90 degrees is depicted in Fig.~\ref{fig:twistedWG}.

We want the length of the twist-element ($l_{\mathrm{twist}}$) to be as short as possible as the electric field inside points in an unfavorable direction. Its minimum length is bound by several times the free space wavelength given by $\lambda_f = 2 \pi c / \omega_{101}$.
In a first approximation the geometric form factor of this structure would then be given by Eq.~\eqref{equ:cForTwisty}, where $l_{\mathrm{straight}}$ is the length of a straight section.

\begin{equation}
\label{equ:cForTwisty}
C \approx 0.66 \frac{l_{\mathrm{straight}}}{l_{\mathrm{straight}}+l_{\mathrm{twist}}}
\end{equation}

For such a structure, operated in a dipole magnet, its sensitivity and resonant frequency would be roughly comparable to a higher order mode cavity operated in a wiggler-type magnet, as described in Sect.~\ref{ch:wiggler}. The maximum total length of the structure $l_{\mathrm{total}}$ is limited by overlapping modes and can be estimated by Eq.~\eqref{maxlength2} if we substitute the magnetic domain length $d$ by $l_{\mathrm{straight}}+l_{\mathrm{twist}}$.

Advantageous for this configuration is that dipole bending magnets from particle accelerators can be used efficiently, they are usually more common and provide higher magnetic field strengths compared to wiggler magnets. Its disadvantage is, that only the straight sections contribute to the active axion detection volume.

\section{Toroidal Magnet}\label{toroid}
Another interesting class of magnets are so-called toroidal magnets as they are currently used in particle physics detectors such as ATLAS.
Design studies~\cite{Irastorza:2011gs} for a next generation axion helioscope (NGAH) also suggest to use such a magnet.
In certain regions, these magnets provide transverse (to the long side) magnetic fields similar to dipole magnets.
The advantage of these magnets is that the corresponding ``bore'' sizes are significantly 
larger than in the case of accelerator dipole magnets.
For example in the magnet configuration worked out in~\cite{Irastorza:2011gs} one (of six available) bores has a radius of about $45\,{\rm cm}$ with a (more or less) dipole like field of $5\,{\rm T}$ over a length of 20 m~\footnote{Indeed, the construction of such a large magnetic configuration for an ALP DM search can be significantly simplified compared to one for solar axions, since neither tracking nor extracting and focusing of X-rays is required.}.

In addition to using a full toroid which provides multiple sections of strong dipole-like fields one could also use a single coil of such a magnet.
A convenient (because already existing) example is the B0 coil build as a test setup for the ATLAS toroid. It provides a 4T field in a 
large region (the coil has dimensions $5{\rm m}\times 9{\rm m}$. Although this field may not be fully homogeneous this should not reduce
the geometric factor~\eqref{equ:geomFactor} dramatically.\

There are multiple ways to use this increased magnetized volume. One is to insert multiple identical cavities into the bore
and couple them together. Using cavities of the same type as in the previous sections would correspond to an increase in volume by a 
factor of a ${\rm few}\times100$ which by itself would allow for a dramatic increase in scanning speed without requiring further improvements.
However, as already mentioned previously, coupling such a large number of cavities is certainly challenging.
Note that with such a configuration one automatically avoids any of the coherence problems discussed at the end of Sect.~\ref{sectioned}.
Moreover, as also mentioned in Sect.~\ref{sectioned}, already an increase in the volume by a factor of 10-20 would help significantly with the scanning speed and correspondingly with the sensitivity.

Alternatively one could use the enormous available magnetic volume to complement the ADMX search at very low frequencies.
Let us take the NGAH magnet as an example (for the B0 coil similar or perhaps even bigger volumes may be possible). 
Using the same ratio $w/h=3/2$ as for the dipole magnets we can now fit a cavity with dimensions $w=75\,{\rm cm}$ 
and $h=50\,{\rm cm}$ with a length
of $20\,{\rm m}$.
This corresponds to a frequency
\begin{equation}
\omega_{101}=(2\pi)200\,{\rm MHz}=8.3\times 10^{-7}\,{\rm eV}.
\end{equation}

Due to the increase in volume to a whopping $7500~\ell$ and the lower axion mass one achieves already a scanning speed of 
$\sim 5\,{\rm THz}/{\rm year}$ with a $Q$-factor of $10^3$ and a noise temperature of 3K at a coupling $g_{a\gamma\gamma}=10^{-15}\,{\rm GeV}^{-1}$.
In other words, even with this fairly conservative setup one can achieve sensitivities $g_{a\gamma\gamma}=10^{-16}\,{\rm GeV}^{-1}$
or better.
Indeed, in addition to the increase in volume and the decrease in the axion mass, the situation is also helped by the fact that lower frequencies typically
allow for better $Q$-factors. Even the quantum limit for the noise temperature is much lower $\sim 5$~mK perhaps even allowing for better
detection. This makes this avenue very promising indeed!

To illustrate this we indicate in Fig.~\ref{limits} the sensitivity for a set of 5 cavities covering the $200-500$~MHz range. For simplicity we have kept the ratio $w/h=3/2$ fixed and scaled the volume appropriately. We have also kept the $Q$-factor fixed. For the green region we have used
$Q=10^3$ and $T_{n}=3\,$K whereas for the red region we have used $Q=10^5$ and $T_{n}=0.03\,$K.
Already with the conservative setup one probes deep into the QCD axion region and with the more ambitious setup the whole range of
couplings predicted for the QCD axion can be covered.

\section{High Field Magnets - Future Prospects}\label{future}
Searches for axions and ALPs will most certainly benefit from
ongoing research and development in the area of high field
magnets that exists in other communities.  
As can be seen, for
example, in Eq.~\eqref{power} the signal power relevant for haloscope searches increase
quadratically with external magnetic field strength. Equation~\eqref{power} also shows that
extra volume also increases the output power but, as we have seen, also gives us the chance to explore
new frequency ranges. Therefore these are the features we will look for in future magnets.  

The High Energy Physics community has interest in high magnetic
field development for its core physics research program at
colliders such as the Large Hadron Collider (LHC) at CERN.
The United States (US) LHC Accelerator Research Program (LARP) is a collaboration 
of four US National Laboratories (Fermilab, BNL, LBNL and SLAC), has been formed to work with CERN to
advance the performance of the LHC by doing R\&D towards magnet
upgrades. Similar goals are being pursued by the Cern High Field Magnet Program and the High-Field Magnet group within the EuCARD consortium.
The development goals include both quadrupoles and dipoles.  The community believes that a vigorous program to develop Nb3Sn and
Nb3Al technology will be required to support this goal~\cite{Zlobin:2003sm,Mokhov:2003ae,Meyers:2011,Rijk1,Rijk2}.
In dipole development, the current record field is
of 20 T with a 10 mm bore at 1.8 K.  However, development goals do not only include high fields, but also large apertures $\gtrsim 100\,{\rm mm}$.
High temperature superconductors
(HTS) for use in high field magnets is also being pursued by these
collaborations. Although HTS is still rather far from being suitable
for use in a more application-oriented program its continued progress suggest that it may eventually be useful at
least in specialized applications such as in this axion and ALP
community.

Accelerators also use quadrupole magnets. Development of quadrupoles with the largest possible
aperture (of order 100 mm) and an operating gradient greater than $200\, {\rm T}/{\rm m}$ is underway. However, as long as their maximal field strengths do not exceed those of dipole magnets they are less suitable for ALP DM searches.

Advanced light sources are at present one of the fastest developing technologies. Therefore wiggler and undulator magnets are also likely to make significant progress in the near future. Already at present there is a significant number of different magnets of this type available (see, e.g.~\cite{Zolotarev} for an overview).
With the aim of achieving more compact aparatuses and shorter wavelengths, higher field strength are one of the main development goals.

The Nuclear Magnetic Resonance research community is broad and
active, with high field magnet development a priority due to
its demanding spectroscopy needs. 
The researchers make use of superconducting as well as resistive magnets.  
(Although pulsed magnets for use in this area are developed and can achieve
incredible performance such as 100 T magnetic fields, this
technology is not easily suited for halo ALP searches that
are described in this document.)  
Most of these magnets are of the solenoid type used in previous ALP searches such as ADMX.
A 45 T hybrid magnet has
already been developed and employed in NMR research.  It was
designed as a versatile, reliable, user-friendly magnet system
capable of producing 45~T in a 32~mm bore. The system consists
of two sets of coils: the superconducting outsert and the
resistive insert. The outsert contains three concentric
cable-in-conduit coils operating at 1.8 K.  It uses 33 MW of
power and is estimated to cost about \$15 million~\cite{largemagnet}.

While developments outside of this present field require specialized
performance parameters (high field uniformity and operation in high
radiation fields, for example) that are not needed in axion and ALP
searches, these magnets should be commercially available in the very
near future and provide attractive alternatives to current magnet
technology.

\section{Conclusions}\label{conclusions}
The last few years have seen significant progress in the development of magnets for accelerators, particle detectors as well as magnets used for the
generation of synchrotron radiation and free electron laser light.
Beyond using them for their intended purpose they may also offer great opportunities for the next generation of searches for one of the favorite
dark matter candidates, axion(-like particle) dark matter (ALP dark matter).

In this note we have investigated a variety of such magnets, in particular accelerator dipoles, wiggler magnets used for light generation,
as well as toroidal magnets used in particle detectors. 
From the solenoidal magnets previously used in ALP dark matter searches they mainly differ in two aspects. First they provide dipole-like fields, second they offer a significantly different geometry. 
In particular the latter allows us to explore new frequency (i.e. mass) ranges which are untouched by previous experiments. 
Both dipole and wiggler magnets have relatively small cross sectional area but an enormous length. When operated in low lying 
TE modes this allows us to reach high frequencies in the range of $5-10$~GHz corresponding to ALP masses of ${\rm few}\times10^{-5}\,{\rm eV}$
which are so far unexplored. At the same time their large length provides a sufficient volume to test very small values for the ALP-photon coupling,
reaching all the way down to the values predicted for the QCD axion. This makes them very promising for searches in this regime.
On the other hand toroidal magnets offer a quite significant cross sectional area as well as enormous length. This could be utilized either by combining multiple long thin cavities as suggested for the dipoles and wigglers, or alternatively to test very low frequencies. The former clearly increases the sensitivities to even smaller couplings. The latter allows to explore an entirely new frequency range 
$\sim{\rm few}\times 100\,{\rm MHz}$ with impressive sensitivity covering the whole range of couplings predicted for the QCD axion.

An important aspect for all ALP dark matter searches is their ability to scan over a variety of ALP masses. This can be achieved by tuning the frequency of the cavity. The different geometry required by the use of dipole magnets also requires a somewhat different tuning technique. Here, we have investigated moveable dielectric inserts which allow to tune the cavity frequency by up to $20\%$ while still offering excellent sensitivity for ALPs.
(Both the quality factor as well as the geometric factor quantifying the coupling of the modes to the ALP field remain high.)

All in all the various magnets investigated in this note offer excellent opportunities to search for ALP dark matter in a wide range of frequencies which are complementary to existing searches with solenoids. Many of these magnets are already available at accelerator and light source facilities
are just waiting to be used to uncover the mystery of dark matter.

\section*{Acknowledgements}
M. Betz and F. Caspers would like to thank the CERN BE-department management, in particular E. Jensen and R. Jones for support.
P. Sikivie would like to thank the CERN Theory Group and the organizers of the DMYH11 Workshop for their hospitality and support.

\end{document}